\documentclass[prd,onecolumn]{revtex4}
\usepackage{dcolumn}
\usepackage{multirow}
\usepackage{graphicx}
\usepackage{amssymb}
\usepackage{bm}
\usepackage{hyperref}
\usepackage{epstopdf}
\usepackage{color}
\usepackage{mathrsfs}
\usepackage{amsmath,amssymb,amsthm}
\usepackage{rotating}
\usepackage{sverb, longtable}
\usepackage{subfigure}

\usepackage[graphicx]{realboxes}
\usepackage{adjustbox}
\begin{document}

\title{JWST high redshift galaxy observations have a strong tension with Planck CMB measurements}
\author{Deng Wang}
\email{cstar@nao.cas.cn}
\author{Yizhou Liu}
\affiliation{National Astronomical Observatories, Chinese Academy of Sciences, Beijing, 100012, China}
\begin{abstract}
JWST high redshift galaxy observations predict a higher star formation efficiency than the standard cosmology does, which poses a new tension to $\Lambda$CDM. We find that the situation is worse than expected. The true situation is that the Planck CMB measurement has a strong tension with JWST high redshift galaxy observations. Specifically, we make a trial to alleviate this tension by considering alternative cosmological models including dark matter-baryon interaction, $f(R)$ gravity and dynamical dark energy. Within current cosmological constraints from Planck-2018 CMB data, we find that these models all fail to explain such a large tension. A possible scenario to escape from cosmological constraints is the extended Press-Schechter formalism, where we consider the local environmental effect on the early formation of massive galaxies. Interestingly, we find that an appropriate value of nonlinear environmental overdensity of a high redshift halo can well explain this tension.    

\end{abstract}
\maketitle

\section{Introduction}
Since the cosmic acceleration is discovered by Type Ia supernovae (SNe Ia) \cite{SupernovaSearchTeam:1998fmf, SupernovaCosmologyProject:1998vns} and confirmed by two independent probes cosmic microwave background (CMB) \cite{WMAP:2003elm, Planck:2013pxb,Planck:2018vyg} and baryon acoustic oscillations (BAO) \cite{Blake:2003rh,Seo:2003}, the standard 6-parameter cosmological model, $\Lambda$-cold dark matter ($\Lambda$CDM) has achieved great success in characterizing the physical phenomena across multiple scales at the background and perturbation levels. However, the validity of $\Lambda$CDM is challenged by various kinds of new observations for a long time, and consequently new puzzles emerge such as the so-called Hubble constant ($H_0$) tension (see \cite{DiValentino:2020vhf,Abdalla:2022yfr} for recent reviews). It is noteworthy that, so far, we can not study effectively the correctness of $\Lambda$CDM around redshift $z\sim10$, since currently mainstream probes BAO and SNe Ia can not give direct observations at high redshifts. The lack of stable high redshift observations will prevent us from testing $\Lambda$CDM more completely during the early stage of the evolution of our universe. 

Very excitingly, the recent released high redshift galaxy observations \cite{Labbe:2022,Harikane:2022,Naidu:2022,Castellano:2022} in the range $z\in[7,11]$ by JWST, which contains a population of surprisingly massive galaxy candidates with stellar masses of order of $10^9 M_\odot$, can help explore whether $\Lambda$CDM is valid at high redshifts. In the literature, Refs.\cite{Labbe:2022,Harikane:2022,Boylan-Kolchin:2022kae,Lovell:2022bhx} have reported the cumulative stellar mass density (CSMD) estimated from early JWST data is higher than that predicted by $\Lambda$CDM within $z\in[7,11]$. Ref.\cite{Menci:2022wia} points out that dynamical dark energy (DDE) can explain this anomalous signal and the corresponding constraint on DDE is displayed. Subsequently, if the nature of dark matter (DM) is fuzzy, this high SMD can be recovered \cite{Gong:2022qjx}. Furthermore, Ref.\cite{Biagetti:2022ode} discusses under which circumstances primordial non-Gaussianity can act as a solution. 

Since these high redshift galaxy observations from JWST have important implications on cosmology and astrophysics, we attempt to probe whether early JWST data indicates any possible signal of new physics. Specifically, we study three classes of beyond $\Lambda$CDM cosmological models, i.e., DM-baryon interaction (DMBI), modified gravity (MG) and DDE. In addition, we consider the case of the extended halo mass function (HMF). We find that Within current cosmological constraints from Planck-2018 CMB obervations, these three models all fail to explain this large tension. A possibly successful scenario to escape from cosmological constraints is the extended Press-Schechter formalism.        

This study is outlined in the following manner. In the next section, we introduce the basic formula of CSMD. In Section III, we review briefly the alternative cosmological models and extended Press-Schechter HMF. In Section IV, numerical results are displayed. The discussions and conclusions are presented in the final section.  

\section{Basic formula}
As shown in Ref.\cite{Labbe:2022}, the  CSMD from early JWST data has a large excess relative to that predicted by $\Lambda$CDM. To explain this excess, we shall briefly introduce the basic formula of the cumulative SMD. The HMF for a given cosmological model reads as    
\begin{equation}
\frac{{\rm d}n}{{\rm d}M} = F(\nu)\frac{\rho_m}{M^2}\left|\frac{{\rm d}\ln\sigma}{{\rm d}\ln M}\right|, \label{1}
\end{equation}
where the function $F(\nu)$ for the Press-Schechter HMF \cite{Press:1973iz} is expressed as 
\begin{equation}
F(\nu)=\sqrt{\frac{2}{\pi}}\nu \mathrm{e}^{-\frac{\nu^2}{2}}, \label{2}
\end{equation}
and $\rho_m$ denotes the average background matter density, $M$ the halo mass, $\sigma$ the variance of smoothed linear matter density field and reads as 
\begin{equation}
\sigma^2(R)=\frac{1}{2\pi^2}\int_{0}^{\infty}k^2P(k)W^2(kR){\rm dk}, \label{3}
\end{equation}
where $k$ is the comoving wavenumber, $P(k)$ the matter power spectrum, $W(kR)=3(\sin kR-kR\cos kR)/(kR)^3$ the Fourier transformation of a spherical top-hat filter with radius $R=[3M/(4\pi\bar\rho_0)]^{1/3}$, $\nu=\delta_c/[D(z)\,\sigma]$ \cite{Carroll:1991mt} ($\delta_c=1.686$ is the critical collapsed density) and $D(z)=g(z)/[g(0)(1+z)]$ the linear growth factor for a specific cosmological model, where $g(z)$ for $\Lambda$CDM reads as
\begin{equation}
g(z)=\frac{5}{2}\Omega_m(z)\left\{\Omega_m(z)^{\frac{4}{7}}-\Omega_\Lambda(z)+\left[1+\frac{\Omega_m(z)}{2}\right]\left[1+\frac{\Omega_\Lambda(z)}{70}\right]\right\}^{-1}, \label{4}
\end{equation}
where $\Omega_m(z)$ and $\Omega_\Lambda(z)$ are energy densities of matter and dark energy (DE) at a given redshift, respectively. 

An effective quantity to study the validity of the $\Lambda$CDM model is the CSMD $\rho_\star$, which can be characterized by a fraction of baryon mass contained within a given DM halo above a certain mass scale $M_\star$ and reads as 
\begin{equation}
\rho_\star(>M_\star,z)=\epsilon f_b\int_{z_1}^{z_2}\int_{\frac{M_\star}{\epsilon f_b}}^{\infty}\frac{{\rm d}n}{{\rm d}M}M{\rm d}M\frac{{\rm d}V}{{\rm d}z}\frac{{\rm d}z}{V(z_1,z_2)}, \label{5}
\end{equation}
where $\epsilon$ is the star formation efficiency, $f_b$ the baryon fraction and $V(z_1,z_2)$ the comoving volume in the redshift range $z\in[z_1,\,z_2]$.

\section{Alternative models}

\subsection{Dark matter-baryon interaction}
Up to now, the standard cosmological paradigm indicates that DM is cold, collisionless and only participates in gravitational interactions \cite{Abdalla:2022yfr}. In light of the lack of experimental detections of DM and emergent cosmological tensions in recent years, the scenario beyond the standard DM assumption becomes more and more attractive. An interesting category is interactions between DM and the Standard Model particles such as baryons, photons and neutrinos. In this study, we consider the case of DMBI. 

The interaction between DM and baryons produces a momentum exchange proportional to momentum transfer cross section, which can be shown as 
\begin{equation}
\sigma_T=\int(1-\cos\theta){\rm d}\Omega\frac{{\rm d}\bar\sigma}{{\rm d}\Omega}, \label{6}
\end{equation}
In the weakly coupled theory, $\sigma_T$ can just depend on even powers of DM-baryon relative velocity $v$ and, in general, it is a power law function of $v$. Here we adopt $\sigma_T=\sigma_{\rm DM-b}v^{n_b}$ and denote the DMBI cross section as $\sigma_{\rm DM-b}$. Specifically, we study the mini-charged DM (DM particle with a fractional electric charge) corresponding to the case of $n_b=-4$, which has been used to explain the anomalous 21 cm signal from EDGES \cite{Barkana:2018lgd}.   

For this model, we introduce two basic assumptions: (i) DM and baryons obey the Maxwell velocity distribution; (ii) both species are non-relativistic. As a consequence, the Euler equation of DM can obtain an extra term $\Gamma_{\rm DM-b}(\theta_b-\theta_{\rm DM})$, where $\Gamma_{\rm DM-b}$ is the conformal DM-baryon momentum exchange rate, and $\theta_{\rm DM}$ and $\theta_b$ represent the velocities of DM and baryons, respectively. At leading order, $\Gamma_{\rm DM-b}$ is expressed in terms of DM bulk velocity and reads as \cite{Dvorkin:2013cea} 
\begin{equation}
\Gamma_{\rm DM-b}=\frac{a\rho_bf_{\rm He}\sigma_{\rm DM-b}c_{-4}}{m_{\rm DM}+m_b}\left(\frac{T_{\rm DM}}{m_{\rm DM}}+\frac{T_b}{m_b}+\frac{V_{\rm RMS}^2}{3}\right)^{-1.5}, \label{7}
\end{equation}
where $a$ is the scale factor, $\rho_b$ the average baryon energy density, $f_{\rm He}\simeq0.76$, $c_{-4}=0.27$ the integration constant (see \cite{Dvorkin:2013cea,Xu:2018efh} for details), and $T_i$ and $m_i$ denote the temperature and average mass of species $i$, respectively. The bulk velocity dispersion can be shown as \cite{Becker:2020hzj}
\begin{equation}
V_{\rm RMS}^2=\left\{\begin{aligned}
&10^{-8}, & z>10^3 \\
&\frac{(1+z)^2}{10}, & z\leq 10^3 \\
\end{aligned}
\right.. \label{8}
\end{equation}   
The interaction between DM and baryons can produce the energy and momentum exchange. It is clear that DMBI reduces to $\Lambda$CDM when $\sigma_{\rm DM-b}=0$. There is a possibility that DMBI can increase the baryon fraction and consequently give a large star formation efficiency. This indicates that DMBI can act as a potential solution to the recent puzzle from JWST data.

\subsection{Modified gravity}
Since general relativity (GR) can not explain current cosmic expansion in the absence of cosmological constant, the modifications in the gravity sector on cosmic scales has inspired a broad interest in order to describe this anomalous phenomenon. Here we shall consider the simplest extension to GR, $f(R)$ gravity, where the modification is a function of Ricci scalar $R$. $f(R)$ gravity was firstly introduced by Buchdahl \cite{Buchdahl:1970zz} in 1970 and more detailed information can be found in recent reviews \cite{DeFelice:2010aj,Sotiriou:2008rp}. Its action is written as
\begin{equation}
S=\int d^4x\sqrt{-g}\left[\frac{f(R)}{2}+\mathcal{L}_m\right], \label{9}
\end{equation}
where $\mathcal{L}_m$ and $g$ denote the matter Lagrangian and the trace of a given metric, respectively. 

For the late-time universe, a viable $f(R)$ gravity scenario should explain the cosmic expansion, pass the local gravity test and satisfy the stability conditions. To investigate whether MG can explain the high redshift galaxy data from JWST, in this study, we consider the so-called Hu-Sawicki $f(R)$ model (hereafter HS model) \cite{Hu:2007nk}, which is characterized by 
\begin{equation}
f(R)=R-\frac{2\Lambda R^{\bar n}}{R^{\bar n}+\mu^{2\bar n}}, \label{10}
\end{equation}  
where $\bar n$ and $\mu$ are two free parameters characterizing this model. By taking $R\gg\mu^2$, the approximate $f(R)$ function can be expressed as 
\begin{equation}
f(R)=R-2\Lambda-\frac{f_{R0}}{\bar n}\frac{R_0^{\bar n+1}}{R^{\bar n}}, \label{11}
\end{equation}
where $R_0$ is the present-day value of Ricci scalar and $f_{R0}=-2\Lambda\mu^2/R_0^2$. Note that HS $f(R)$ gravity reduces to $\Lambda$CDM when $f_{R0}=0$.  

An intriguing question is whether recent JWST anomaly is a signal of beyond GR. We will carefully analyze this possibility in this study.

\subsection{Dynamical dark energy}
Although Ref.\cite{Menci:2022wia} has claimed that DDE can explain the large CSMD from JWST, we think their method is inappropriate and consequently their result maybe incorrect. We need to reanalyze the case of DDE.   

As is well known, the equation of state (EoS) of DE $w=-1$ in the standard cosmological model. However, starting from observations, the doubt about the correctness of $\Lambda$CDM stimulates the community to explore whether DE is dynamical over time or not. In general, one depicts the DDE model by a simple Taylor expansion of DE EoS, i.e., $\omega(a)=\omega_0+(1-a)\omega_a$ \cite{Chevallier:2000qy,Linder:2002et}, where $\omega_a$ characterizes the time evolution of DE EoS. The dimensionless Hubble parameter is expressed as 
\begin{equation}
E_{\mathrm{DDE}}(z)=\left[\Omega_{m}(1+z)^3+(1-\Omega_{m})(1+z)^{3(1+\omega_0+\omega_a)\mathrm{e}^{\frac{-3\omega_az}{1+z}}}\right]^{\frac{1}{2}}. \label{12}
\end{equation}
Note that this model is a two-parameter extension to $\Lambda$CDM and it reduces to $\Lambda$CDM when $\omega_0=-1$ and $\omega_a=0$. 

\subsection{Extended halo mass function}
When applied into a complicated gravity system, the function of Press-Schechter HMF is limited, since it does not consider the nonlinear environmental effects. To overcome this shortcoming, the extended Press-Schechter (EPS) HMF is proposed in Ref.\cite{Bond:1990iw} and reads as 
\begin{equation}
\frac{{\rm d}n(M_1,z|M_2,\delta_2)}{{\rm d}M_1}=\frac{M_2}{M_1}f_m(S_1,\delta_1|S_2,\delta_2)\left|\frac{{\rm d}S_1}{{\rm d}M_1}\right|, \label{13}
\end{equation} 
where the mass variance $S_1=\sigma^2(M_1)$ and $S_2=\sigma^2(M_2)$ (see Eq.(\ref{3})), and one can obtain the average number of progenitors at time $t_1$ in the mass range $(M_1, M_1+{\rm d}M_1)$ which by time $t_2$ ($t_2>t_1$) have merged to form a large halo of mass $M_2$. The multiplicity function $f_m$ is expressed as 
\begin{equation}
f_m(S_1,\delta_1|S_2,\delta_2)=\frac{1}{\sqrt{2\pi}}\frac{\delta_1-\delta_2}{(S_1-S_2)^{3/2}}{\rm exp}\left[-\frac{(\delta_1-\delta_2)^2}{2(S_1-S_2)}\right]dS_1. \label{14}
\end{equation} 
$\delta_1$ and $\delta_2$ are, respectively, the linear overdensities in spherical regions of masses $M_1$ and $M_2$. To study the environmental impacts on the high redshift HMF, we choose $M_2$ as a present-day halo corresponding to current overdensity $\delta_2$. To compute $\delta_2$, one should transform the nonlinear overdensity $\delta_{nl}$ at redshift z in Eulerian space into the linear overdensity in Lagrangian space. The corresponding analytic fitting formula based on spherical collapse model is \cite{Mo:1995cs,Sheth:2001dp}
\begin{equation}
\delta_2(\delta_{nl},z)=\frac{\delta_1}{1.68647}\left[1.68647-\frac{1.35}{(1+\delta_{nl})^{2/3}}-\frac{1.12431}{(1+\delta_{nl})^{1/2}}+\frac{0.78785}{(1+\delta_{nl})^{0.58661}}\right]. \label{15}
\end{equation} 
Since there is a possibility that the excessively high CSMD from JWST is caused by nonlinear environmental effect, we attempt to explain it using the EPS formalism.

\begin{figure}
	\centering
	\includegraphics[scale=0.55]{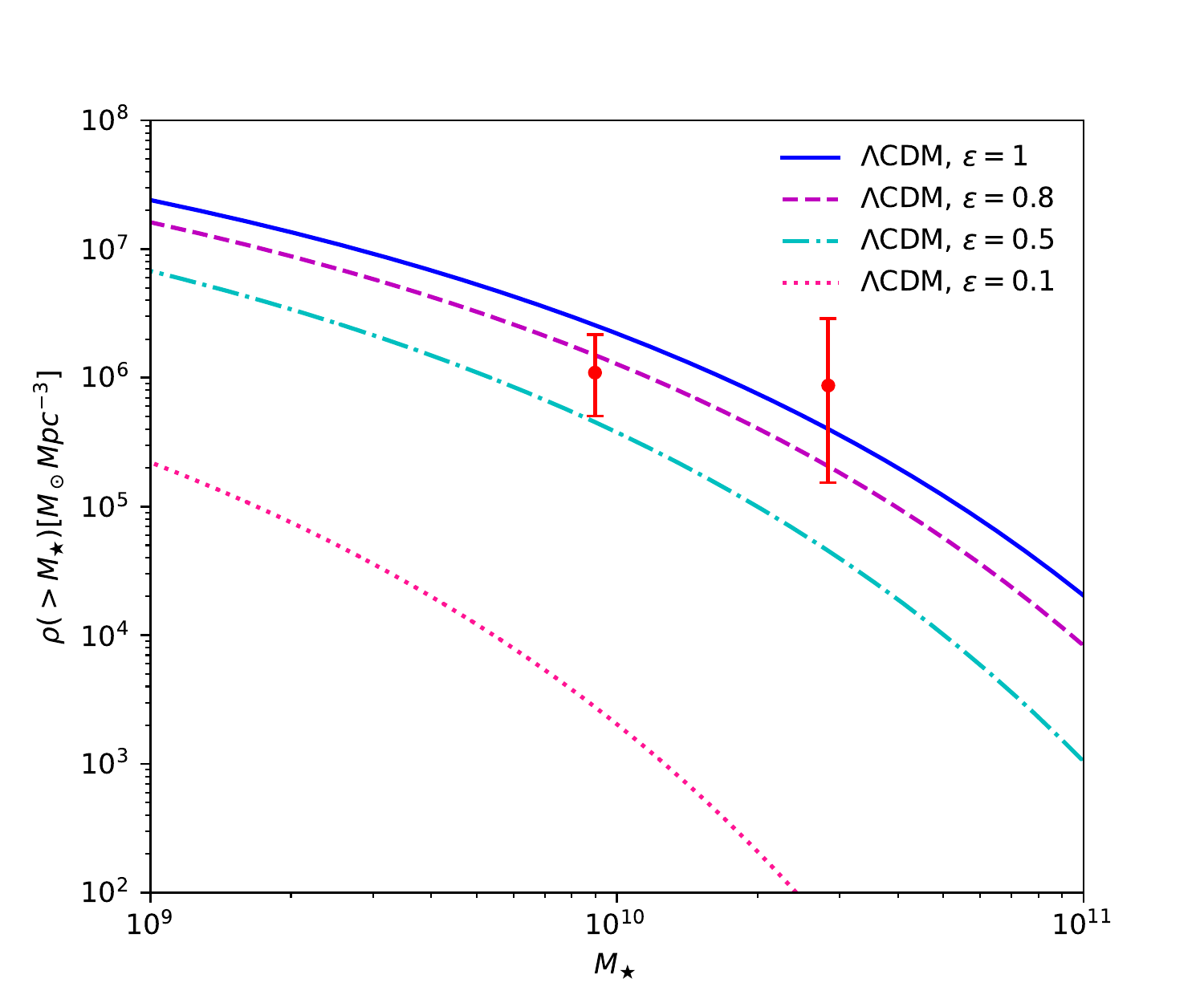}
	\includegraphics[scale=0.55]{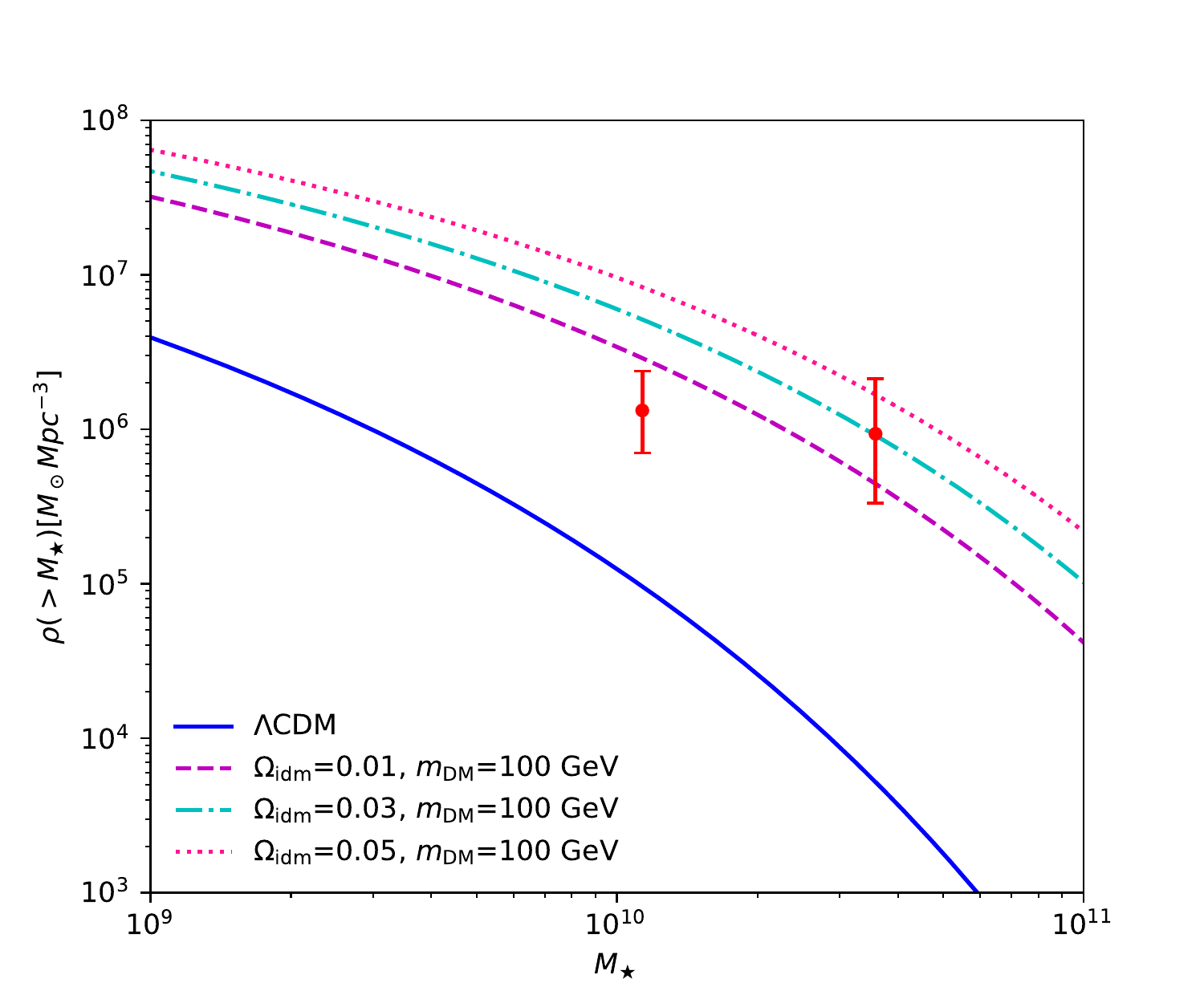}
	\includegraphics[scale=0.55]{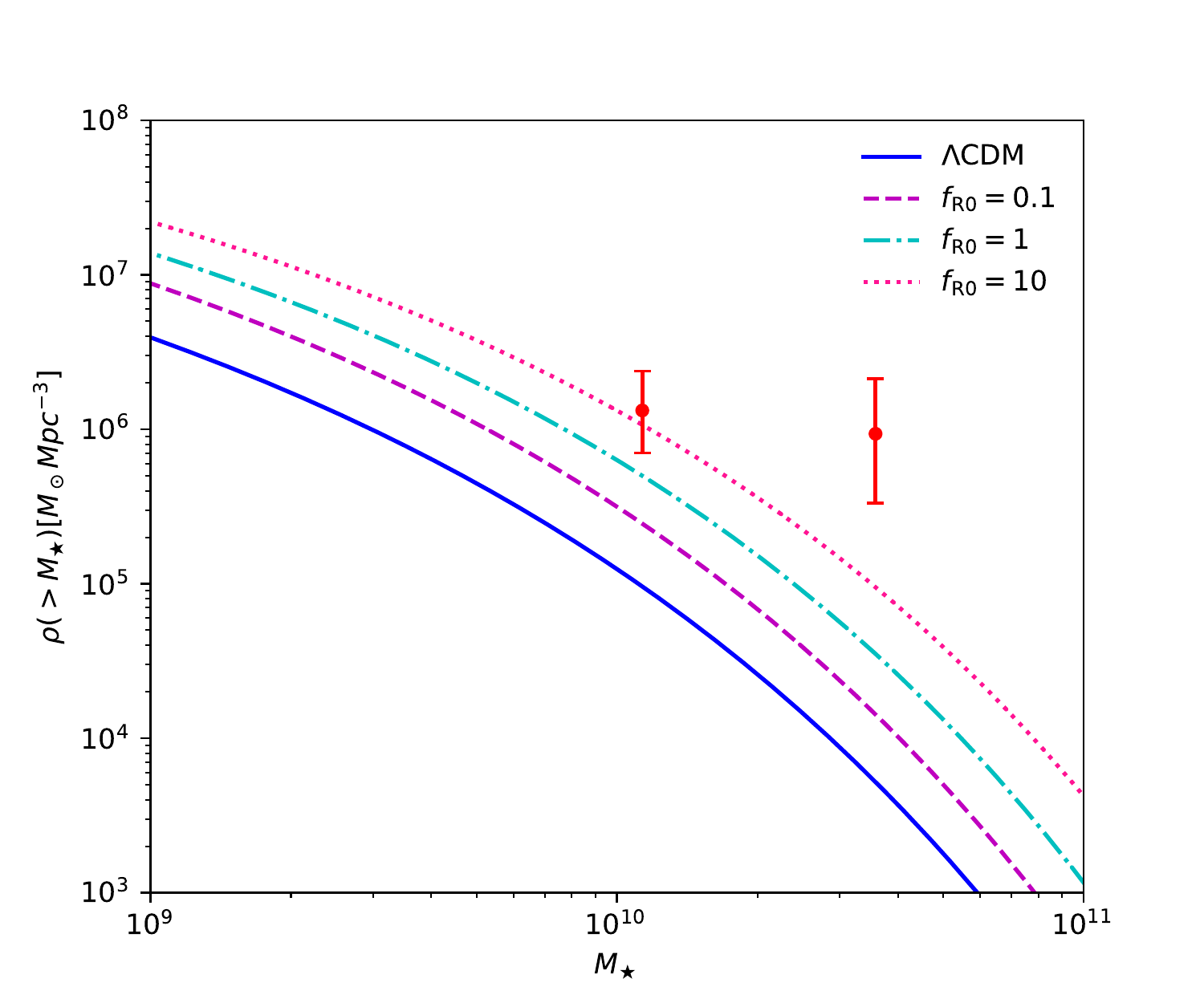}
	\includegraphics[scale=0.55]{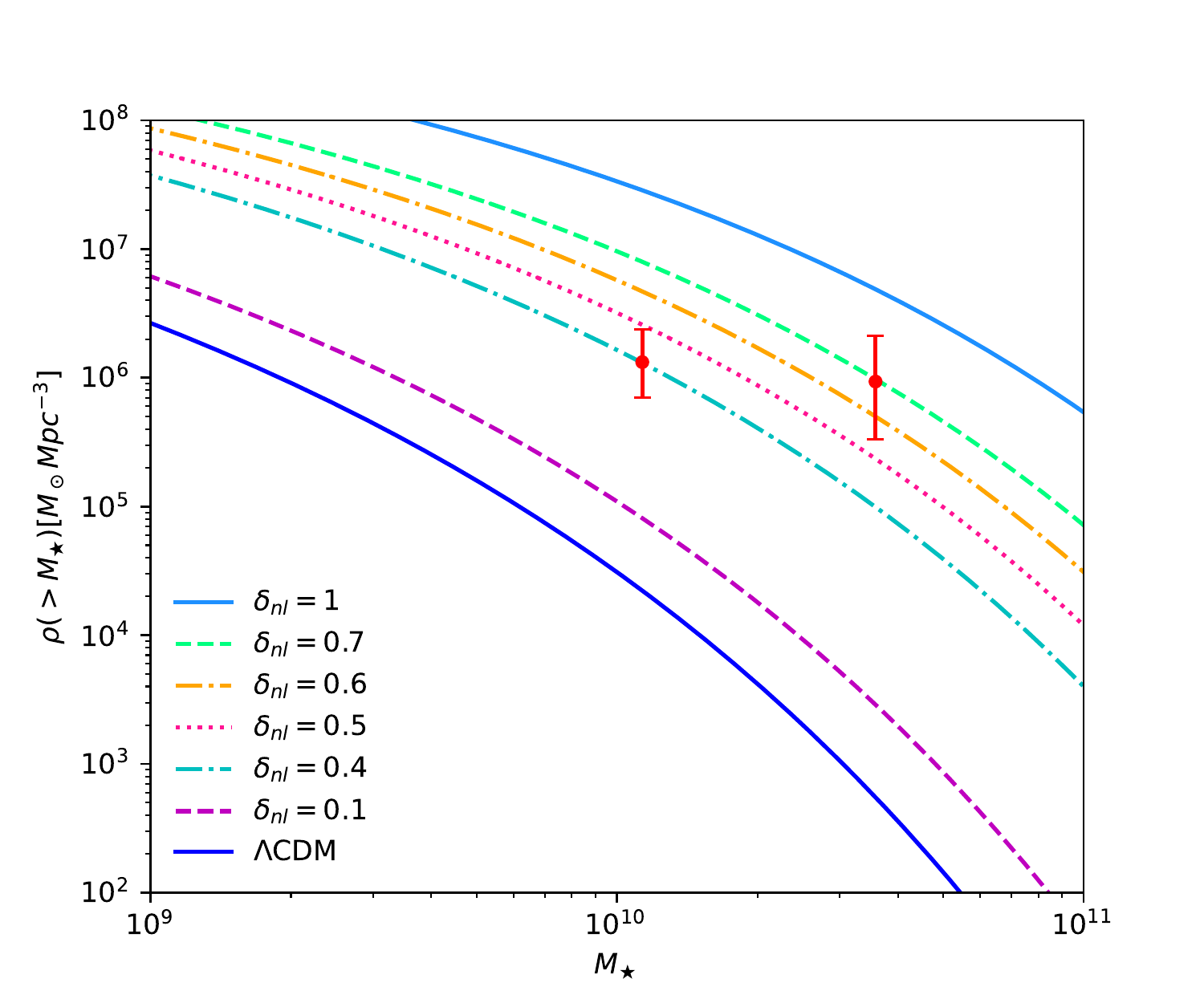}
	\caption{The CSMDs for the $\Lambda$CDM, DMBI, $f(R)$ gravity and EPS models are shown from top to bottom and left to right, respectively. Note that for $\Lambda$CDM, we compute the CSMDs in the redshift range $z\in[7,9]$ by choosing different values of the SFE $\epsilon$. For the other models, we calculate the CSMDs in the redshift range $z\in[9,11]$ when $\epsilon=1$.}\label{f1}
\end{figure}

\begin{figure}
	\centering
	\includegraphics[scale=0.3]{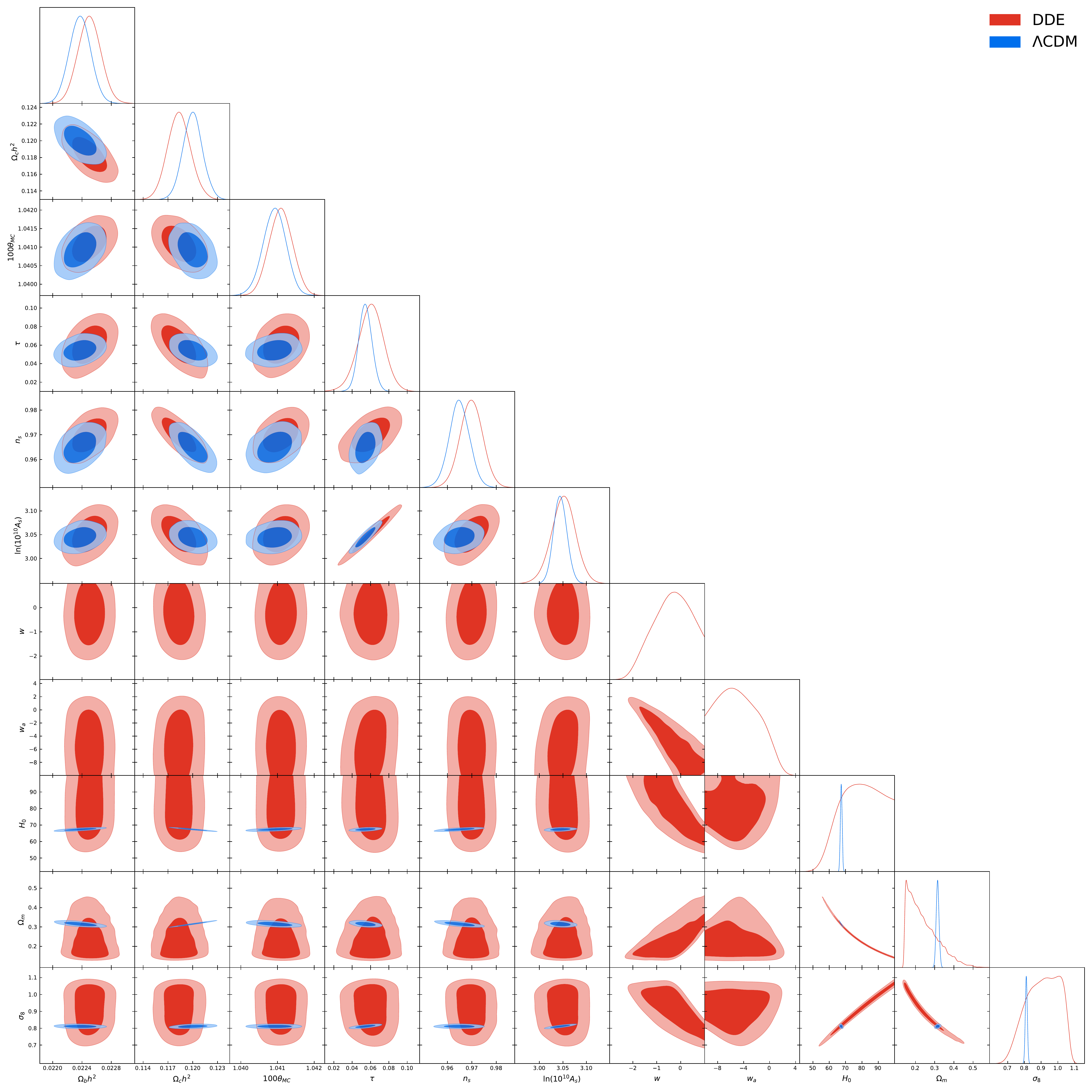}
	\caption{The marginalized posterior probability distributions of the $\Lambda$CDM and DDE models from the Planck-2018 CMB constraints are shown.}\label{f2}
\end{figure}

\begin{figure}
	\centering
	\includegraphics[scale=0.385]{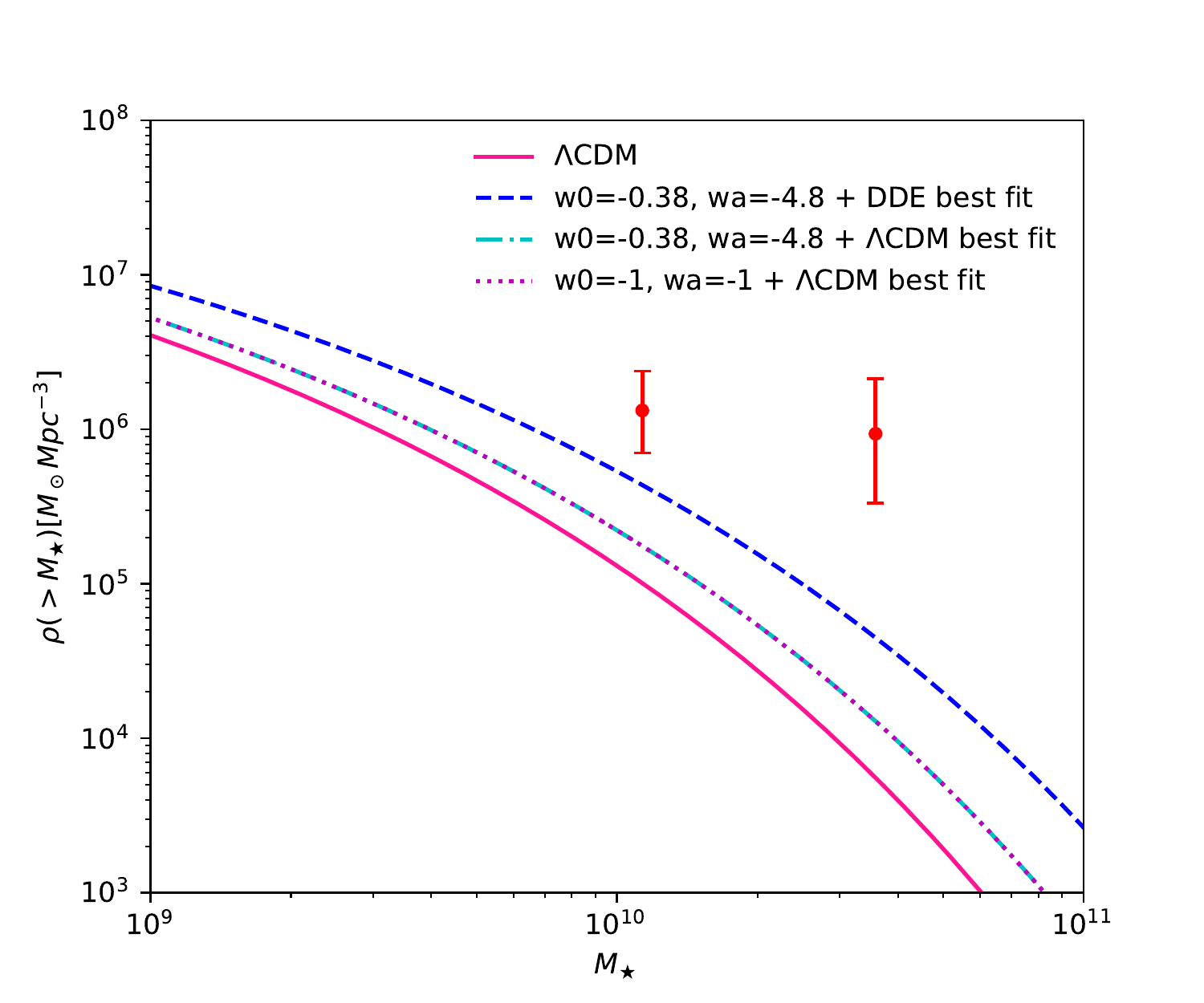}
	\includegraphics[scale=0.385]{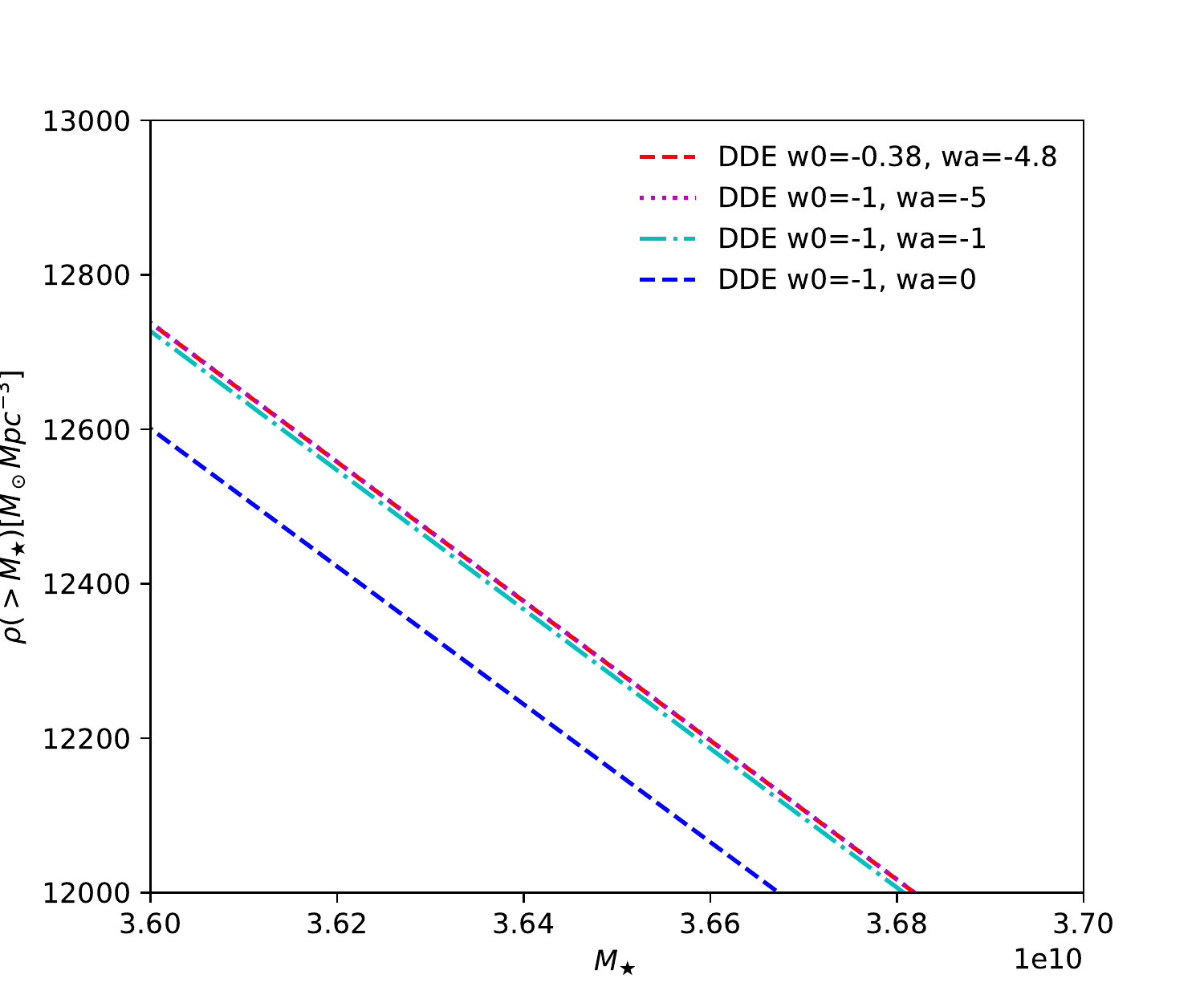}
	\includegraphics[scale=0.385]{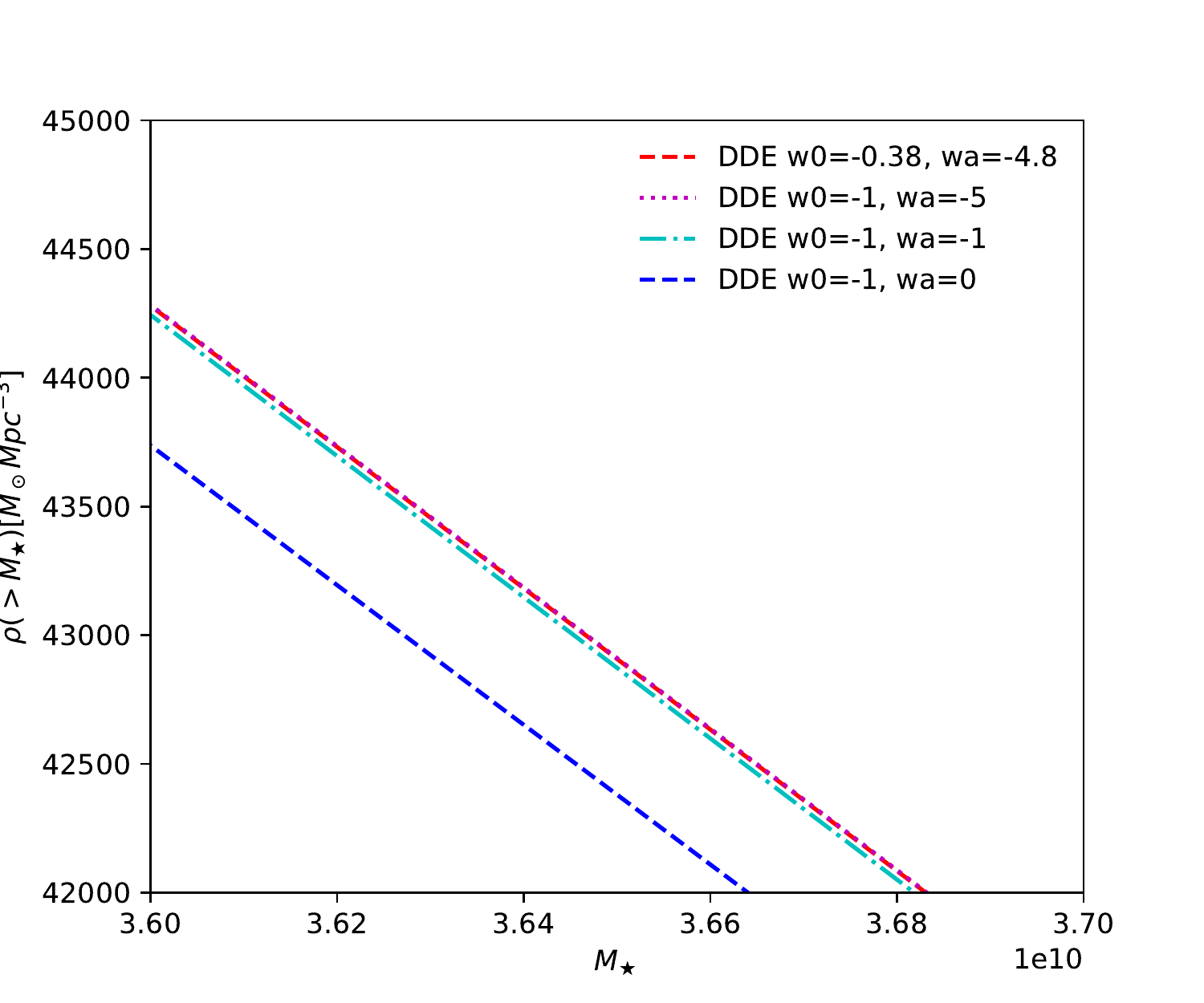}
	\caption{The CSMDs of the DDE model computed at in the redshift range $z\in[9,11]$ are shown when assuming $\epsilon=1$. {\it Left}: Different combinations of parameter values and best fits from constraints, respectively. {\it Medium}: Only the $\Lambda$CDM best fit; {\it Right}: Only the DDE best fit.}\label{f3}
\end{figure}

\section{Methods and results}
At first, we employ the best fits from current cosmological constraints as our baseline values for four models. Since we hope that the following calculations can be permitted by present-day observations, our discussions and results will mainly focus on the allowed parameter space. Then, for different models, we use different Boltzmann codes to calculate their background evolution, growth factors and matter power spectrum at different redshifts. Specifically, we take \texttt{CLASS} \cite{Dvorkin:2013cea,Slatyer:2018aqg,Xu:2018efh} for DMBI and use modified \texttt{CAMB} \cite{Lewis:1999bs,Lewis:2013hha} for $f(R)$ gravity, DDE and EPS scenarios. Note that $\Lambda$CDM is adopted in the EPS scenario. Subsequently, we compute the HMF at different redshifts for the above four models. Finally, we work out the maximal CSMD for each model according to the permitted parameter space, and check whether these scenarios are consistent with the latest JWST data. Notice that Eq.(\ref{4}) is only used in the EPS model and the growth factors of the other three models are obtained from the corresponding software package.

Our numerical analysis results are presented in Figs.\ref{f1}-\ref{f3}. At first, we display the CSMD of $\Lambda$CDM in the redshift range $z\in[7,9]$ and see its performance. In general, the SFE $\epsilon$ is about $10\%$ according to current observational constraints \cite{Harikane:2022}. Nonetheless, one can see that in the top left panel of Fig.\ref{f1}, $10\%$ is nowhere near enough to reach the lower bounds of JWST data points in $\Lambda$CDM. One needs the star formation rate in galaxies to be at least $50\%$ in order to explain the inconsistency. In the meanwhile, one can easily find that $\epsilon=0.8$ can successfully explain two data points but $100\%$ SFE can not. Except for $\Lambda$CDM, we all calculate the maximal CSMD in the other models, i.e., assuming $\epsilon=1$. 

In the second place, we make a trial to explore whether alternative cosmological models can alleviate even solve the tension between JWST and Planck CMB observations. In the DMBI case, we attempt to acquire a higher baryon fraction by the coupling between DM and baryons, and consequently explain this discrepancy occurred in $\Lambda$CDM. However, we find that varying coupling strength $\sigma_{\rm DM-b}$ hardly affects the CSMD, and only the variation of interaction DM fraction $\Omega_{\rm idm}$ affects significantly the CSMD. When assuming the DM particle mass $m_{\rm DM}=100$ GeV, the cross section $\sigma_{\rm DM-b}=10^{-42} \, \mathrm{cm}^2$ and choosing the fraction $\Omega_{\rm idm}=0.01$, 0.03 and 0.05, this tension can be efficiently relieved but it seems that this model is difficult to explain both data points. However, if considering the current cosmological constraint that gives a very small $\Omega_{\rm idm}$ \cite{Becker:2020hzj}, DMBI still behaves like $\Lambda$CDM and can not resolve this discrepancy. We have also studied the impacts of $m_{\rm DM}$ and find different DM particle masses also can not explain JWST data. 

In $f(R)$ gravity, we find small $f_{R0}$ such as 0.1 and 1 can not expalin the anomaly but a very large value $f_{R0}=10$ can do. This implies that one needs a large deviation from GR to be responsible for JWST data. Unfortunately, the latest cosmological constraint gives $\log_{10} f_{R0}< -6.32$ at the $2\,\sigma$ confidence level \cite{Wang:2022xdw}, which is much smaller than 10. Therefore, similar to DMBI, $f(R)$ gravity also fails to alleviate this tension. Interestingly, this gives us a hint that, if two galaxies observed by JWST are located in the low density region of the universe where MG effect is very large, the data can be appropriately explained. 

Furthermore, we are interested in whether the nature an simple extension to $\Lambda$CDM, DDE, can explain the inconsistency. As mentioned above, Ref.\cite{Menci:2022wia} claimed that JWST data can clearly constrain DDE. However, within current constraining precision, we query this conclusion. To ensure the validity of our conclusion, we constrain $\Lambda$CDM and DDE models using the Planck-2018 CMB temperature and polarization data (see Fig.\ref{f2}), and then obtain the best fitting values of parameters of these two models. One can easily find the constrained values of model parameters of $\Lambda$CDM in Ref.\cite{Planck:2018vyg}. For DDE, we obtain current baryon and CDM densities $\Omega_bh^2=0.0225$ and $\Omega_ch^2=0.1184$, the ratio between angular diameter distance and sound horizon at the redshift of last scattering $\theta_{MC}=1.04109$, the optical depth due to the reionization $\tau=0.06$, the amplitude and spectral index of primordial power spectrum $A_s=2.114\times10^{-9}$ and $n_s=0.9698$, and two DE EoS parameters $\omega_0=-0.38$ and $\omega_a=-4.8$. Same as DMBI and $f(R)$ gravity models, we use the same method to work out the CSMD of DDE, and find that the variation of the CSMD is largely dominated by the values of six basic parameters $\Omega_bh^2$, $\Omega_ch^2$, $\theta_{MC}$, $\tau$, $A_s$ and $n_s$. Although $\omega_0$ and $\omega_a$ is loosely constrained by CMB data (constrained $\omega_0$-$\omega_a$ parameter space is large), different values of $\omega_0$ and $\omega_a$ hardly affect the CSMD. For instance, in the left panel of Fig.\ref{f3}, $\omega_0=-0.38$ and $\omega_a=-4.8$ plus the $\Lambda$CDM and DDE best fits gives completely different CSMDs. Choosing the $\Lambda$CDM best fit, $(\omega_0,\omega_a)=(-0.38,-4.8)$ and $(\omega_0,\omega_a)=(-1,-1)$ gives very similar results in the logarithmic space. In the medium and right panels of Fig.\ref{3}, we verifies that taking same best fits of $\Lambda$CDM and DDE, respectively, choosing different DE EoS parameter pair just produces very limited differences. After scanning the DDE parameter space, we find clearly that DDE also can not explain this tension, but its best fit can help increase the value of CSMD and become closer to JWST data points (see the left panel of Fig.\ref{3}). The reason that the result in Ref.\cite{Menci:2022wia} is different from ours is that they do not implement an appropriate cosmological constraint based on the Planck CMB data.            

The result from $f(R)$ gravity prompts us to study the environmental effect of JWST galaxies on the CSMD. The most straightforward method is replacing the Press-Schechter HMF with the EPS formalism in the framework of $\Lambda$CDM, where the sole parameter $\delta_{nl}$ characterizes the nonlinear environmental effect of a high redshift halo. In the bottom right panel, we calculate the maximal CSMDs in the redshift range $z\in[9,11]$ for the EPS model. We find that neither overlarge ($\delta_{nl}=1$) nor too small ($\delta_{nl}=0.1$) explain JWST observations and that the larger $\delta_{nl}$ is, the larger the CSMD is. Since the total sky area covered by the JWST initial observation is large enough ($\sim40$ armin$^2$) \cite{Labbe:2022}, we can not rule out this possibly local environmental effect. However, unfortunately, there is no $\delta_{nl}$ passing two data points simultaneously.

\section{Discussions and conclusions}
Recently, the early data release of JWST reveals the possible existence of high redshift galaxies. What is interesting is these galaxies in the redshift range $z\in[7,11]$ exhibit the overlarge star formation rate, which is incompatible with the prediction of the standard cosmology. This may indicate that JWST data contain the signal of new physics.

In this study, we try to resolve this tension with alternative cosmological models including DMBI, $f(R)$ gravity and DDE. We find that in light of the precision of current cosmological constraint from Planck-2018 CMB data, these models all fail to explain this large tension. Specifically, for DMBI, the coupling strength $\sigma_{\rm DM-b}$ between DM and baryons hardly affects the CSMD. For $f(R)$ gravity, the effect of varying $f_{R0}$ on the CSMD is too small to relive the tension. For DDE, although the constrained DE EoS parameter space is large, different parameter pair $(\omega_0,\omega_a)$ just produces very limited differences in the CSMD. Interestingly, a large interacting DM fraction and a large deviation from Einstein's gravity can both generate a large CSMD. 

A possible scenario to escape from current cosmological constraints is the EPS formalism, where we consider the local environmental effect on the CSMD. We find that an appropriate value of nonlinear environmental overdensity of a high redshift halo can well explain the CSMD discrepancy. However, we do not find an EPS model that can simultaneously explain two data points.

In the near future, JWST will bring more useful data to human beings, so that we can extract more physical information to uncover the mysterious veil of nature.

\section*{Acknowledgments}
DW warmly thanks Liang Gao, Jie Wang and Qi Guo for helpful discussions. We thank Hang Yang for letting us notice the JWST related works. This study is supported by the National Nature Science Foundation of China under Grants No.11988101 and No.11851301.

\end{document}